\documentclass[
reprint,
 amsmath,amssymb,
 aps,
prl,
tightenlines
]{revtex4-1}
\usepackage[colorlinks,urlcolor=blue, citecolor=blue,linkcolor=blue]
{hyperref}
\usepackage{natbib} 
\usepackage{hyperref} 
\usepackage{siunitx}
\usepackage{booktabs}
\usepackage{color}
\usepackage{multirow}
\usepackage{overpic}
\usepackage{colortbl}
\usepackage{graphicx}
\usepackage{float}
\usepackage{array}
\usepackage{makecell}
\usepackage{subfigure}
\usepackage{graphicx}
\usepackage{dcolumn}
\usepackage{bm}
\usepackage{xcolor}
\usepackage{amsmath}
\usepackage{natbib}
\usepackage{comment}

\definecolor{aa}{RGB}{0,0,139}

\newcommand{\EE}{e^{+}e^{-}}

\newcommand{\jpsi}{J/\psi}

\newcommand{\psipp}{\psi(3770)}
\newcommand{\MM}{\mu^+\mu^-}

\usepackage[mathlines]{lineno}
\usepackage{multirow}
\usepackage{enumerate}
\usepackage{amsmath}
\newcommand{\BESIIIorcid}[1]{\href{https://orcid.org/#1}{\hspace*{0.1em}\raisebox{-0.45ex}{\includegraphics[width=1em]{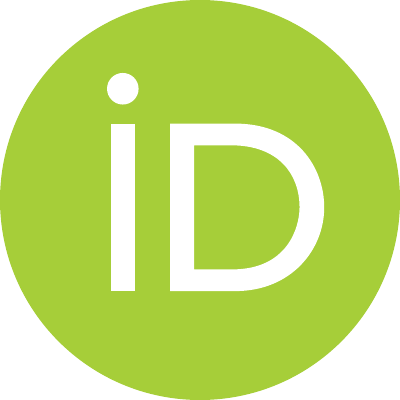}}}} 

\begin{document}
\title{\boldmath Evidence of $\psipp \to \pi^{0}\jpsi$ }

\author{
\begin{small}
\begin{center}
M.~Ablikim$^{1}$\BESIIIorcid{0000-0002-3935-619X}, M.~N.~Achasov$^{4,c}$\BESIIIorcid{0000-0002-9400-8622},
P.~Adlarson$^{82}$\BESIIIorcid{0000-0001-6280-3851},
X.~C.~Ai$^{88}$\BESIIIorcid{0000-0003-3856-2415},
C.~S.~Akondi$^{31A,31B}$\BESIIIorcid{0000-0001-6303-5217},
R.~Aliberti$^{39}$\BESIIIorcid{0000-0003-3500-4012},
A.~Amoroso$^{81A,81C}$\BESIIIorcid{0000-0002-3095-8610},
Q.~An$^{78,65,\dagger}$,
Y.~H.~An$^{88}$\BESIIIorcid{0009-0008-3419-0849},
M.~S.~Anderson$^{39}$\BESIIIorcid{0009-0008-1550-2632}
% ...
\\
\vspace{0.2cm}
(BESIII Collaboration)\\
\vspace{0.2cm} {\it
$^{1}$ Institute of High Energy Physics, Beijing 100049, People's Republic of China\\
% ...
$^{4}$ Budker Institute of Nuclear Physics SB RAS (BINP), Novosibirsk 630090, Russia\\
% ...
$^{31}$ INFN Sezione di Ferrara, (A)INFN Sezione di Ferrara, I-44122, Ferrara, Italy; (B)University of Ferrara, I-44122, Ferrara, Italy\\
% ...
$^{39}$ Johannes Gutenberg University of Mainz, Johann-Joachim-Becher-Weg 45, D-55099 Mainz, Germany\\
% ...
$^{65}$ State Key Laboratory of Particle Detection and Electronics, Beijing 100049, Hefei 230026, People's Republic of China\\
% ...
$^{78}$ University of Science and Technology of China, Hefei 230026, People's Republic of China\\
% ...
$^{81}$ University of Turin and INFN, (A)University of Turin, I-10125, Turin, Italy; (B)University of Eastern Piedmont, I-15121, Alessandria, Italy; (C)INFN, I-10125, Turin, Italy\\
$^{82}$ Uppsala University, Box 516, SE-75120 Uppsala, Sweden\\
% ...
$^{88}$ Zhengzhou University, Zhengzhou 450001, People's Republic of China\\
% ...
$^{c}$ Also at the Novosibirsk State University, Novosibirsk, 630090, Russia\\
% ...
}
\end{center}
\vspace{0.4cm}
\end{small}
}

\begin{abstract}
We report the first evidence for the hadronic transition $\psi(3770) \to \pi^0 J/\psi$ using a data sample of $20.3~\mathrm{fb}^{-1}$ collected at $\sqrt{s} = 3.773$~GeV with the BESIII detector. The $e^+ e^- \to \pi^{0}J/\psi$ process is observed with a statistical significance of $6.1\sigma$, while the significance specifically attributed to $\psi(3770) \to \pi^{0}J/\psi$ is $4.5\sigma$. We measure the dressed cross section for $e^+e^- \to \pi^0 J/\psi$ to be $(249 \pm 44 \pm 15)~\text{fb}$ and determine the branching fraction $\mathcal{B}(\psi(3770) \to \pi^0 J/\psi) = (2.08 \pm 0.36 \pm 0.21\pm 0.25) \times 10^{-5}$, where the first uncertainty is statistical, the second is systematic, and the third due to a possible interference with the $\psi(3686) \to \pi^{0}J/\psi$ decay. This is the first determination of this branching fraction. It lies significantly below tetraquark model predictions but aligns with calculations that incorporate meson loop effects, providing crucial insight into the isospin-violating nature of $\psi(3770)$ decays.
\end{abstract}
\maketitle
\oddsidemargin  -0.2cm
\evensidemargin -0.2cm

Discovered in the 1970s following the $J/\psi$ and $\psi(3686)$, the $\psi(3770)$ is the lowest-mass charmonium resonance lying above the $D\bar{D}$ threshold~\cite{psppdiscover}. While the charmonium model predicts that the $1^{3}D_{1}$ state should decay predominantly ($>99\%$) into open-charm final states~\cite{pure1D}, the observation of $\psi(3770) \to \pi^{+}\pi^{-} J/\psi$ in 2003, with a branching fraction of $(0.34 \pm 0.14 \pm 0.09)$\%, challenged this prediction~\cite{psppfirstobserve}. Subsequent measurements by BESII and BESIII further revealed that non-$D\bar{D}$ decays account for a significant fraction, $(15.1 \pm 5.6 \pm 1.8)$\%, of the total hadronic width~\cite{nondd1,nondd2,nondd3,nondd4,ParticleDataGroup:2008zun}. However, this sizable inclusive rate stands in stark contrast to the sum of observed exclusive non-$D\bar{D}$ branching fractions, which remains below 2\%~\cite{PDG}. 
A recent K-matrix analysis suggests an upper limit of $6$\% for non-$D\bar{D}$ decays at 90\% confidence level~\cite{nonDD-BF2}. The tension amongst these results underscores the pressing need for further investigation into exclusive decay channels to resolve this apparent discrepancy.

The nature of the $\psi(3770)$ is central to several open questions in charmonium physics. Theoretical interpretations include a $2^{3}S_{1}$--$1^{3}D_{1}$ mixing scheme, initially proposed to resolve the ``$\rho-\pi$" puzzle~\cite{Rosner}, a possible four-quark composition~\cite{Voloshin}, and recent BESIII results indicating that the $\psi(3770)$ may be a multi-resonance complex~\cite{BESIII:591}. In this context, a precise measurement of the hadronic transition $\psi(3770) \to \pi^0 J/\psi$ is important for understanding non-$D\bar{D}$ decays and testing model predictions.
 It was first searched for approximately two decades ago by CLEO. No significant signal was observed, and an upper limit on the branching fraction, $\mathcal{B}(\psi(3770) \to \pi^0 J/\psi) < 2.8 \times 10^{-4}$ at the 90\% confidence level, was established~\cite{CLEOresult}. To date, only two processes yielding $\pi^0 J/\psi$ final states have been established in high-energy experiments: $Z_c(3900) \to \pi^0 J/\psi$~\cite{BESIII:2015cld} and $\psi(3686) \to \pi^0 J/\psi$~\cite{Oreglia:1980zn}. The latter decay is recognized as an isospin-violating process. Isospin conservation is generally an excellent approximation in strong interactions, owing to the small mass difference between the $u/d$ quarks. Consequently, observed violations provide valuable insight into quark mass differences~\cite{Gross:1979ur}, and help refine understanding of mechanisms such as the partial conservation of the axial current and gluon-meson couplings~\cite{Ioffe:1980mx}. 
Determining whether $\psi(3770) \to \pi^0 J/\psi$  conserves or violates isospin is crucial, since this decay could be significantly enhanced if the $\psi(3770)$ contains a non-negligible four-quark component with isospin $I=1$, as suggested in Ref.~\cite{Voloshin}. 
Theoretical predictions for $\mathcal{B}(\psi(3770) \to \pi^0 J/\psi)$ vary widely, from $2.0\times 10^{-4}$ to $4.6\times 10^{-6}$, depending on the model assumptions~\cite{Ioffe:1980mx,Voloshin,predictpi0jpsi2,updatepi0jpsi3,Achasov:2005qb}. Therefore, a precise measurement of this branching fraction provides a critical test on these theoretical assumptions.

Comparing the ratios $\rm{R}^{\Psi} = \frac{\mathcal{B}(\Psi\to\pi^{0}J/\psi)}{\mathcal{B}(\Psi\to\eta J/\psi)}$, where $\Psi$ indicates the $\psi(3686)$ or $\psi(3770)$ mesons, can test the consistency of their decay behaviors and shed light on the internal structure of the $\psi(3770)$. Based on Ref.~\cite{Ioffe:1980mx}, if the $\Psi$ states are pure $c\bar{c}$, the $R^\Psi$ values only depend on the masses of $u/d$ quarks and $\pi/\eta$ mesons, couplings of $\pi/\eta$ with light quark currents, and the phase space. Taking into account the above-mentioned effects, the theoretical calculation yields a ratio ~\mbox{$\rm{R}^{\psi(3686)}/\rm{R}^{\psi(3770)} = 4.0$}.
Recently, BESIII reported a measurement of $\mathcal{B}(\psi(3770) \to \eta J/\psi) = (11.3 \pm 5.9 \pm 1.1)\times 10^{-4}$~\cite{yizhuo}. This result provides a benchmark for such a comparison and enables a similar study with $\psi(3770) \to \pi^0J/\psi$. In this Letter, we present a measurement of this decay channel, compare our result with the previous upper limit and theoretical predictions, and discuss its implications.

This analysis employs the world's largest electron-positron collision data sample at ~\mbox{$\sqrt{s} = 3.773$}~GeV, collected by the BESIII detector at BEPCII, corresponding to an integrated luminosity of ~\mbox{20.3 $\rm{fb^{-1}}$}~\cite{datasample}. In this Letter, we present the dressed cross section for $e^+ e^- \to \pi^0 J/\psi$ at $\sqrt{s} = 3.773$~GeV for the first time. We then extract $\mathcal{B}(\psi(3770) \to \pi^0 J/\psi)$ by accounting for initial state radiation~(ISR) process and the contribution from $e^+ e^- \to \psi(3686) \to \pi^0 J/\psi$. 
The continuum contribution is estimated with the data sample at $\sqrt{s} = 3.650$~GeV with an integrated luminosity of 401.0 pb$^{-1}$~\cite{Luminosity-3650} as well as the data samples at $\sqrt{s}=3.768$~GeV and 3.780 GeV with integrated luminosities of 412.5 pb$^{-1}$ and 407.2 pb$^{-1}$, respectively~\cite{datasample}.

The BESIII detector's specifications and performance are detailed in Ref.~\cite{Ablikim:2009aa}. The response of the experimental apparatus is studied with a detailed {\sc geant4}-based~\cite{geant4} Monte Carlo (MC) simulation of the BESIII detector for particle trajectories, and generated by the generator {\sc kkmc}~\cite{Jadach:2000ir, Jadach:1999vf} with initial state radiation effects~\cite{Kuraev:1985hb} and final state radiation effects~\cite{Richter-Was:1992hxq} included. The decay chain of the signal process $e^+ e^- \to \pi^0 J/\psi$, $J/\psi \to \mu^+ \mu^-$, and $\pi^0 \to \gamma \gamma$ are simulated with the HELAMP (helicity amplitude), VLL (vector lepton lepton), and phase space configurations of {\sc evtgen}~\cite{Lange:2001uf, Ping:2008zz}, respectively. The ISR effect is modeled via the input dressed cross section line shape, where contributions from the \(\psi(3686)\) and \(\psi(3770)\) resonances are included, with parameters fixed to their world-averaged values~\cite{PDG}. The line shape is iteratively determined by updating the branching fraction of $\psi(3770) \to \pi^0 J/\psi$ and the ISR correction factors. The inclusive MC sample includes the production of $D\bar{D}$ pairs, the non-$D\bar{D}$ decays of the $\psi(3770)$, the ISR production of the $J/\psi$ and $\psi(3686)$ states, and the continuum processes incorporated in {\sc kkmc}~\cite{Jadach:2000ir, Jadach:1999vf}. 

Signal events are characterized by two opposite charged tracks and two photons. Charged tracks are required to satisfy $\lvert \cos\theta \rvert < 0.93$ in polar angle, with collision vertices within \(|V_z| < 10\,\text{cm}\) along the beam axis and \(V_{xy} < 1\,\text{cm}\) radially from the interaction point.  Photon candidates are identified as isolated electromagnetic calorimeter (EMC) showers with energy deposits exceeding \(25\,\text{MeV}\) in the barrel region ($\lvert \cos\theta \rvert  < 0.80$) or \(50\,\text{MeV}\) in the end cap $(0.86 < \lvert \cos\theta \rvert < 0.92)$. Showers are excluded if their angular separation from any charged track is within \(10^\circ\) relative to the interaction point, and the EMC timing is constrained to \([0, 700]\,\text{ns}\) relative to the event trigger to suppress noise. Events must contain exactly two charged tracks of opposite charge and at least two photon candidates.

Assuming the charged tracks are muon candidates from a \( J/\psi \) decay, their momenta($p$) are required to satisfy \( 1.2 < p < 1.9~\text{GeV}/c \), with EMC energy deposits \( < 0.4~\text{GeV} \). A vertex fit constraining their origin to the interaction point is applied. To improve signal resolution and suppress backgrounds, a four-constraint (4C) kinematic fit is performed for the process \( e^+e^- \to \gamma\gamma \mu^+\mu^- \), looping over all photon candidates and selecting the pair with the minimum $ \chi^2_{\text{4C}}$. The requirement $ \chi^2_{\text{4C}}<38$ obtained by optimizing the figure-of-merit \( S/\sqrt{S + B} \), where \( S \) denotes the signal yield scaled to preliminary analysis results, and \( B \) represents the background yield from the inclusive MC samples. 

\begin{figure}[htbp]
\centering
\includegraphics[width=0.48\textwidth]{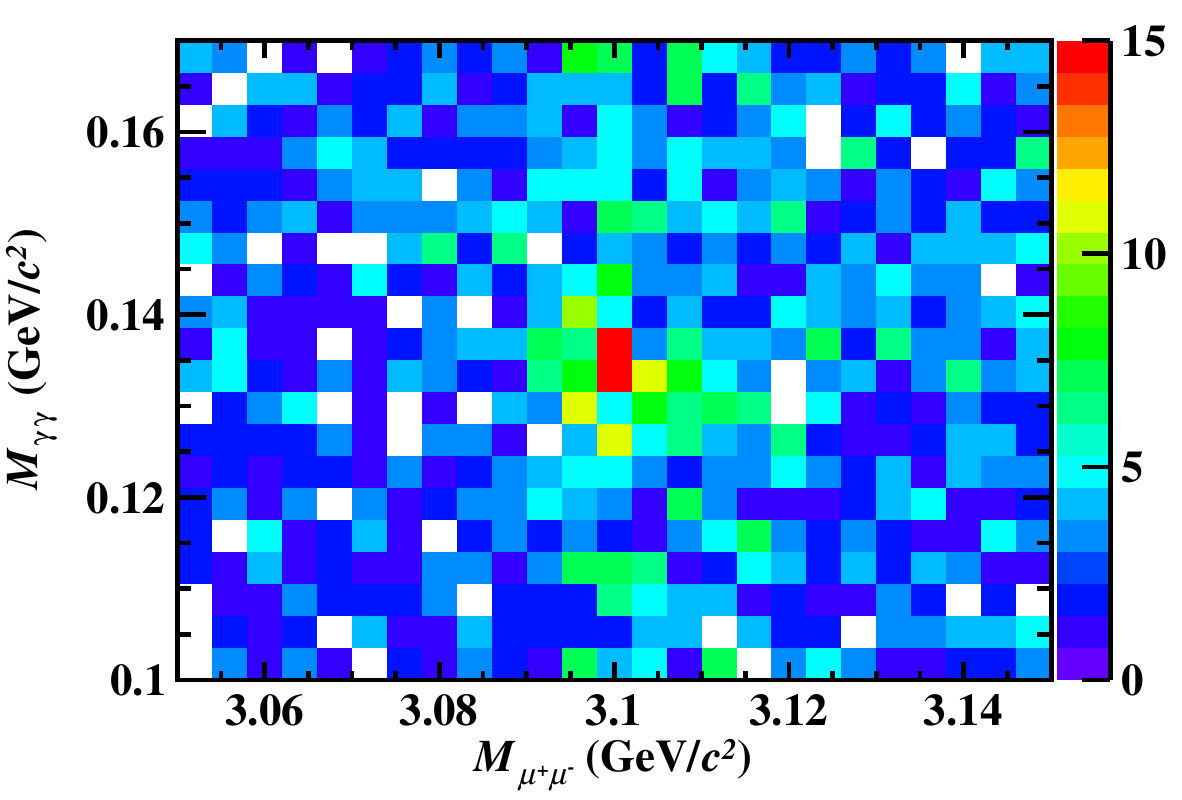}
\caption{Scatter plot of $M_{\mu^+\mu^-}$ versus $M_{\gamma\gamma}$ for the data at $\sqrt{s} = 3.773~$GeV. The color scale indicates the number of events.}
    \label{Fig:ggmumu}
\end{figure}

A study of the inclusive MC sample shows that backgrounds predominantly originate from: (1) $e^+e^- \to \gamma \gamma \mu^+\mu^-$, (2) $e^+e^- \to \gamma^{\rm{ISR}} J/\psi \to \gamma^{\text{ISR}} \mu^+\mu^-$, ~\mbox{(3) $\psi(3770)$} $\to \gamma\chi_{c1} \to \gamma\gamma J/\psi \to \gamma\gamma \mu^+\mu^-$, and ~\mbox{(4) $e^+e^-$} $ \to \pi^+\pi^-\pi^0(\gamma)$. Suppression of these channels is obtained by applying successive criteria: (1) The $\chi^2_{\text{4C}} < \chi^2_{1\gamma}$ requirement ($\chi^2_{1\gamma}$ denoting the $\chi^2$ of the 4C kinematic fit for the $\gamma \mu^{+}\mu^{-}$ hypothesis) eliminates $>90\%$ of di-$\mu$ process, (2) $M_{\gamma_{\text{H}}\mu^+\mu^-} < 3.71~\text{GeV/$c^{2}$}$ ($\gamma_{\text{H}}$ denoting the photon with higher energy) suppresses ISR processes, (3)~the mass window $M_{\gamma_{\text{H}}\mu^+\mu^-} < 3.50~\text{GeV/$c^{2}$}$ or $M_{\gamma_{\text{H}}\mu^+\mu^-} > 3.52~\text{GeV/$c^{2}$}$ vetoes  $\psi(3770) \to \gamma \chi_{c1}$ decays, (4)~muon identification via $\rm{Dep}_{\mu^+} > 40~\text{cm}$ or $\rm{Dep}_{\mu^-} > 40~\text{cm}$ mitigates $e^+e^- \to \pi^{+}\pi^{-}\pi^{0}(\gamma)$ process due to $\pi/\mu$ misidentification, where ``$\rm{Dep}$" represents the tracking hit depth in the muon counter. All selection criteria are optimized by maximizing the figure of merit. Following the application of these criteria, the invariant mass scatter plot of invariant masses ($M_{\mu^+\mu^-}$ versus $M_{\gamma\gamma}$) at $\sqrt{s} = 3.773$~GeV is displayed in Fig.~\ref{Fig:ggmumu}, revealing a distinct $\pi^0 J/\psi$ signal.
 
\begin{figure}[!htbp]
 \begin{overpic}[width=0.48\textwidth]{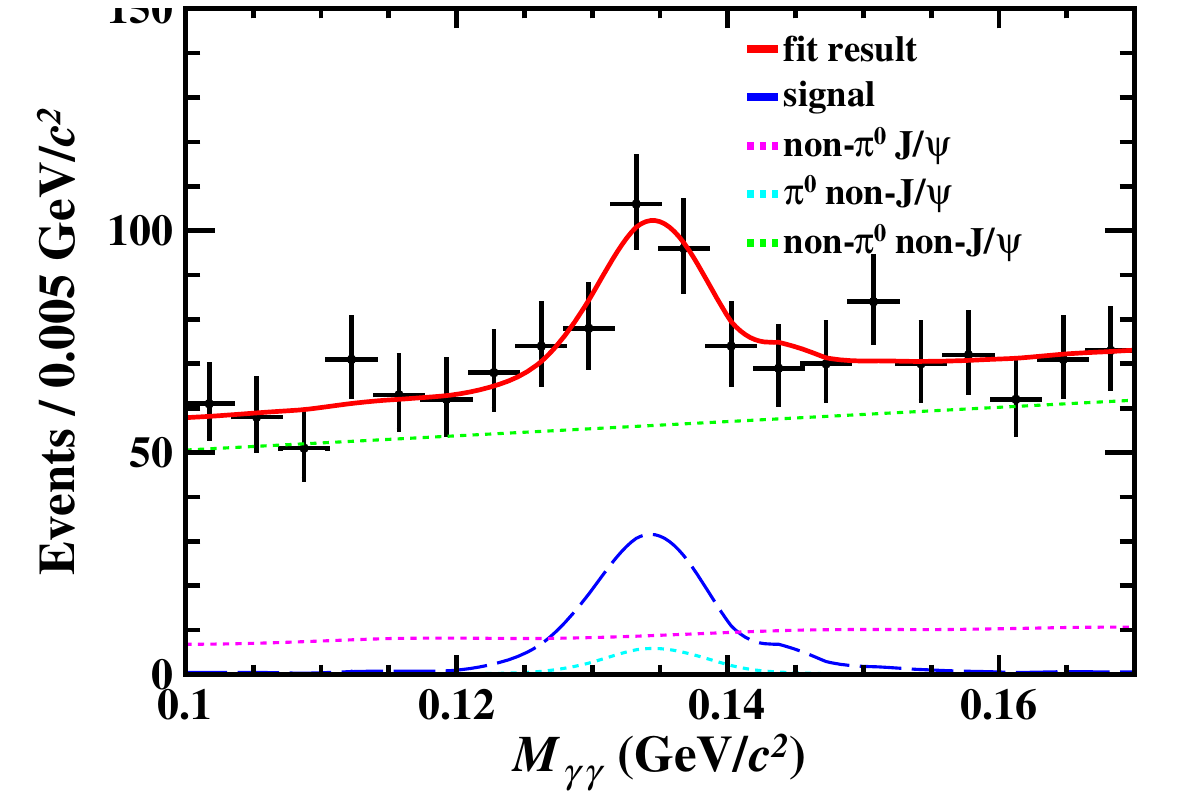}
    \end{overpic}
    \begin{overpic}[width=0.48\textwidth]{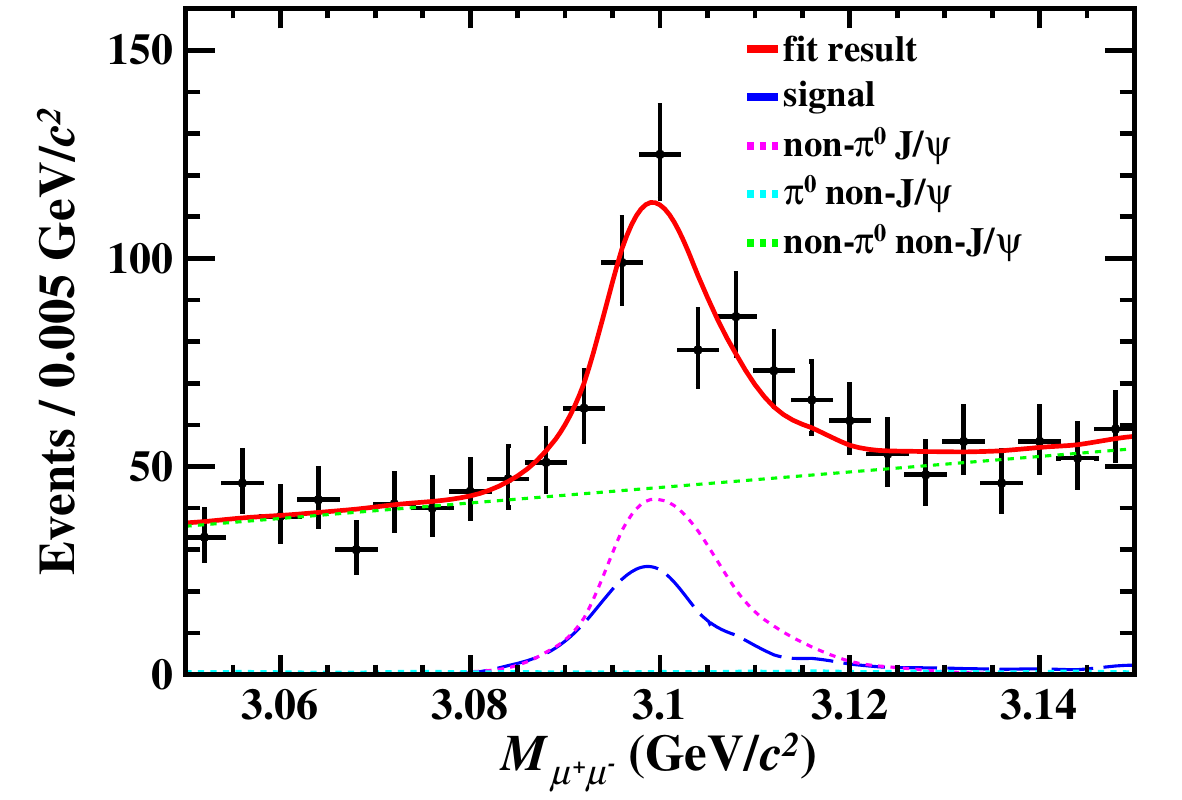}
    \end{overpic}
    \caption{The unbinned two-dimensional likelihood fit to the data at $\sqrt{s}=3.773$~GeV projecting to $M_{\gamma\gamma}$ and $M_{\mu^+\mu^-}$. Dots with error bars represent data, the solid red lines represent the total fit results, the dashed blue lines denote the signal, the dashed pink lines denote the ``non-$\pi^0$ $J/\psi$" backgrounds, the dashed cyan lines denote the “$\pi^0$ non-$J/\psi$” backgrounds, and the dashed green lines denote “non-$\pi^0$ non-$J/\psi$” backgrounds.
}
    \label{Fig:fitresult}
\end{figure}

To extract the signal yield, a two-dimensional unbinned likelihood fit to $M_{\gamma \gamma}$ versus $M_{\MM}$ is performed. The signal shape is extracted from the signal MC simulation. The backgrounds are classified into three categories “$\pi^0$ non-$J/\psi$”, “non-$\pi^0$ $J/\psi$”, and “non-$\pi^0$ non-$J/\psi$”, that are modeled by the shape from the inclusive MC samples, the shape from corresponding exclusive MC samples, and a two-dimensional 1st-order polynomial function with three free parameters, respectively. The yields of each component are determined by the fit. Figure~\ref{Fig:fitresult} shows the fit results to the data taken at $\sqrt{s}=3.773$~GeV, giving the signal yield $N^{\rm{sig}}=114 \pm 20$, where the uncertainty is statistical only. The statistical significance of the signal is determined to be $6.1\sigma$, which is calculated by the difference of the likelihoods and the degrees of freedom with and without the signal component in the fit~\cite{likelihood}.

The continuum contribution from $e^+e^- \to \gamma^{\rm{ISR}}\pi^{0}J/\psi$ is investigated by using the data samples at $\sqrt{s}=3.650, 3.768$, and 3.780 GeV. After applying the selection criteria, no signal with significance $>2\sigma$ is observed. Additionally, no significant signal of direct $e^+ e^- \to \pi^0 J/\psi$ is observed from the data sample taken at $\sqrt{s}>4$~GeV with integrated luminosity of 4832.7~$\rm{pb^{-1}}$~\cite{continuecorss}. Therefore, the measurement of the cross section of $e^{+}e^{-}\to\pi^{0}J/\psi$ is based solely on the contribution from $e^{+}e^{-}\to\psi(3770)\to\pi^{0}J/\psi$  and the tail of high energy side of $e^{+}e^{-} \to \psi(3686) \to \pi^{0}J/\psi$, while potential interference effects between $\psi(3686)$ and $\psi(3770)$ and the direct $e^{+}e^{-}\to \pi^{0}J/\psi$ production are neglected at current  precision.

The dressed cross section of $e^+ e^- \to \pi^{0}J/\psi$ at ~\mbox{$\sqrt{s}=3.773$ GeV} is determined by 
\begin{equation}
\label{eq:cs}
    \sigma^{D}(e^+ e^-  \to\pi^{0}J/\psi) = \frac{N^{\rm{sig}}}{\mathcal{L}\cdot(1+\delta^{\rm{ISR}})\cdot \epsilon \cdot \mathcal{B}_{0}} \ ,
\end{equation}
where $\mathcal{L}$ is the integrated luminosity, $\epsilon$ is the detection efficiency, $\mathcal{B}_{0}$ is the product of branching fractions of $J/\psi\to\MM$ and $\pi^{0} \to \gamma\gamma$ taken from the Particle Data Group (PDG)~\cite{PDG}, and $(1+\delta^{\rm{ISR}})$ is the ISR  factor~\cite{Kuraev:1985hb}. Table~\ref{Tab:informaation} shows the measured dressed cross section at $\sqrt{s}=3.773$ GeV and the values of the other parameters
in Eq.~\ref{eq:cs}. The efficiency and ISR factors are determined iteratively~\cite{Sun:2020ehv}, in which the line shape of the dressed cross sections is parameterized as 
\begin{linenomath*}
\begin{equation}
\label{eq:45}
\sigma^{D}(\sqrt{s}) = \sum_i
\left|
\frac{\sqrt{12\pi\Gamma^{ee}_{\psi_i}\Gamma^{\rm{tot}}_{\psi_i}\mathcal{B}_{\pi^{0}J/\psi}}}{s-M_{\psi_i}^{2}+i\Gamma^{\rm{tot}}_{\psi_i}M_{\psi_i}} \sqrt{\frac {PS(\sqrt{s})}{PS(M_{\psi_i})}}
\right|^{2},
\end{equation}
\end{linenomath*}
where $\psi_i$ represent $\psi(3686)$ or $\psi(3770)$. The mass ($M_{\psi_i}$), total width ($\Gamma^{\rm{tot}}_{\psi_i}$), partial width ($\Gamma^{ee}_{\psi_i}$), and the branching fraction are taken from the PDG~\cite{PDG}. The two-body phase space factor is defined as
\begin{linenomath*}
\begin{equation}
{PS(\sqrt s)=q^{3}(\sqrt s)/s},
\end{equation}
\end{linenomath*}
where $q(\sqrt{s})$ is the momentum of $\pi^0$ in the center-of-mass system at $\sqrt{s}$. Based on Eq.~\ref{eq:cs} and Eq.~\ref{eq:45}, the dressed cross section  $\sigma^D(e^+ e^- \to \psi(3770) \to \pi^0 J/\psi)$ at $\sqrt{s}=3.773$ GeV is calculated by $\sigma^D(e^+ e^- \to \pi^0 J/\psi) - \sigma^D(e^+ e^- \to \psi(3686) \to \pi^0 J/\psi)$~\cite{sigma3686} at $\sqrt{s}=3.773$ GeV, and it is determined to be $(207 \pm 36)$ fb, where the uncertainty is statistical only. The significance of the process is determined to be $4.5\sigma$ after considering both statistical and systematic uncertainties.

\begin{table*}[!htbp]
	\centering
	\caption{The integrated luminosity $\mathcal{L}$, the detection efficiency $\epsilon$, the radiative correction factor $1+\delta^{\rm{ISR}}$, the product of branching fractions of $J/\psi\to\MM$ and $\pi^{0} \to \gamma\gamma$ $\mathcal{B}_{0}$, the signal yield $N^{\rm{sig}}$, and the dressed cross section of $e^+ e^- \to \pi^{0}J/\psi$ at ~\mbox{$\sqrt{s}= 3.773$ GeV}. The uncertainties on the signal yield and the dressed cross section are statistical only. }
	\setlength{\tabcolsep}{1.0mm}{
 	\begin{tabular}{c|c|c|c|c|c}
	\hline \hline
$\mathcal{L}~(\rm {fb^{-1}})$ & $\epsilon~(\%)$ & $(1+\delta^{\rm{ISR}})$ & $\mathcal{B}_{0}$ & $N^{\rm{sig}}$  & $\sigma^{D}(\EE \to \pi^{0} \jpsi)$~(fb)  ~\\	\hline
 20.3& 1.1 & 34.9 & 0.0589 & $114 \pm 20$ & $249 \pm 44$ \\
\hline \hline
	\end{tabular}}
	\label{Tab:informaation}
\end{table*}

The branching fraction of $\psi(3770) \to \pi^0 J/\psi$ is calculated as

\begin{equation}
\small
\label{eq:br}
    \mathcal{B}(\psi(3770) \to \pi^{0}J/\psi) = \frac{\sigma^{D}(\EE \to \psipp \to \pi^{0}J/\psi)}{\sigma^{D}(\EE \to \psi(3770))}\ ,
\end{equation}
where $\sigma^D(\EE \to \psi(3770))$ is the total dressed cross section of the production of $\psi(3770)$ in the electron-positron annihilation that is calculated by using the formula similar to the summed term in Eq.~\ref{eq:45} without the branching fraction factor,  giving $(9.94 \pm 0.77)$ nb with parameters taken from the PDG~\cite{PDG}. This gives a branching fraction for $\mathcal{B}(\psi(3770) \to \pi^0 J/\psi)$ of $(2.08 \pm 0.36 \pm 0.21 \pm 0.25)\times 10^{-5}$, where the second and third uncertainties are the systematic uncertainties, discussed below.  
 
The systematic uncertainties associated with the branching fraction measurement originate from several sources. These include the integrated luminosity, lepton tracking, photon detection, veto criteria, kinematic fit, signal yield extraction, input cross section line shape, the branching fractions $\mathcal{B}(J/\psi \to \mu^{+}\mu^{-})$ and $\mathcal{B}(\pi^{0} \to \gamma\gamma)$, as well as the cited dressed cross sections $\sigma^D(e^+e^- \to \psi(3686) \to \pi^0 J/\psi)$ and $\sigma^D(e^+e^- \to \psi(3770))$. For $\sigma^D(e^+e^- \to \psi(3770) \to \pi^0 J/\psi)$, all the aforementioned sources are considered except the uncertainty related to $\sigma^D(e^+e^- \to \psi(3770))$.
The integrated luminosity, determined via Bhabha scattering events, carries an uncertainty of 0.4\%~\cite{datasample}, which is dominated by the uncertainty of the generator. The systematic uncertainties from tracking are evaluated to be 1.0\% per track, using
the control sample $\psi(3686) \to \pi^{+}\pi^{-}J/\psi$~\cite{sys_track}. The systematic uncertainties from photon reconstruction are estimated to be 1.0\% per photon with the control sample $J/\psi\to\pi^+\pi^-\pi^{0}$~\cite{Photon1,Photon2,Photon3}. The acceptance efficiency of the Dep$_{\mu^\pm}$  selection criterion is evaluated using the control sample $\psi(3770) \to \pi^{0} \pi^{0} J/\psi$. An efficiency correction factor $f = 1.064 \pm 0.004$ is determined. After correcting the MC efficiency to data, we take 0.4\% as the systematic uncertainty.
To estimate the uncertainties associated with the veto conditions on $M_{\gamma_{\text{H}} \mu^+ \mu^-}$, the invariant mass distribution is smeared according to the difference in resolution between data and simulation. The systematic uncertainty is evaluated by comparing the selection efficiencies before and after smearing. For the veto at $M_{\gamma_{\text{H}} \mu^{+} \mu^{-}} < 3.50$~GeV/$c^{2}$ or $ M_{\gamma_{\text{H}} \mu^{+} \mu^{-}} > 3.52$~GeV/$c^{2}$, the resulting uncertainty is 0.3\%. In the case of the $M_{\gamma_{\text{H}} \mu^+ \mu^-} < 3.71$~GeV/$c^{2}$ is the systematic uncertainty 1.0\%.
The systematic uncertainty from the 4C kinematic fit is assessed by correcting track helix parameters in MC simulations. The efficiency difference after correction, 0.4\%, is adopted as the systematic uncertainty. The systematic uncertainty in the signal yield arises from the signal/background parameterizations and fit range. The uncertainty in the signal shape ($3.5\%$) is evaluated by replacing the MC-simulated shape with a double Gaussian function. The uncertainties in the background shapes are evaluated via two aspects: the shapes of ``non-$\pi^0$ $J/\psi$" and ``$\pi^0$ non-$J/\psi$" components, is smeared with the Gaussian function with parameters from $\psi(3686) \to \pi^0 J/\psi$, and the polynomial function of the ``non-$\pi^0$ non-$J/\psi$"  component is changed from first to second order. The resultant deviation, 1.8\%, is taken as the uncertainties for the second term, while the influence from the other two terms is negligible. The Barlow method~\cite{Barlow} is used to test the effect of varying the fitting range. The results show no significant changes within statistical fluctuations, and no further uncertainty is applied. The systematic uncertainty in the input cross section line shape is dominated by the uncertainty in the  $\mathcal{B}(\psi(3770) \to \pi^0 J/\psi)$ measurement, and is evaluated by varying it by $\pm 1\sigma$. The change in $\epsilon\cdot(1+\delta^{\text{ISR}})$, 3.2\%, is assigned as the uncertainty. The $\mathcal{B}(J/\psi \to \mu^{+}\mu^{-})$ and $\mathcal{B}(\pi^{0} \to \gamma\gamma)$ are quoted from the PDG~\cite{PDG}, with uncertainties of $0.6\%$ and $0.1\%$, respectively.
The dressed cross sections $\sigma^D(e^+e^- \to \psi(3686) \to \pi^0 J/\psi)$ and $\sigma^D(e^+e^- \to \psi(3770))$ at $\sqrt{s} = 3.773$~GeV are computed using the resonance parameters (mass, total width, and electron partial widths) of $\psi(3686)$ and $\psi(3770)$ quoted from the PDG~\cite{PDG}. The uncertainties propagated are $0.8\%$ and $7.8\%$ for $\sigma^D(e^+e^- \to \psi(3686)$ and $\sigma^D(e^+e^- \to \psi(3770))$, respectively.

All systematic uncertainties are summarized in Table~\ref{Tab:uncertainty}. The total uncertainty is taken as the quadrature sum of individual contributions, assuming that they are independent with each other. The total uncertainties are $6.0\%$ for $\sigma^D(e^{+}e^{-} \to \psi(3770) \to \pi^{0}J/\psi)$ and $9.9\%$ for $\mathcal{B}(\psi(3770) \to \pi^{0}J/\psi)$, respectively. Potential interference between the $\psi(3686)$ and $\psi(3770)$ resonances is not included in the nominal analysis. A conservative systematic uncertainty is evaluated by taking the maximum cross-section difference between the phase angles $\phi = 0^\circ$ and $\phi = 180^\circ$. This approach accounts for the full extent of an eventual interference effect, yielding an uncertainty of 11.9\%.

\begin{table}[!h]
    \centering
    \setlength{\belowcaptionskip}{10pt}
    \caption{The systematic uncertainties (in \%) for the dressed cross section $\sigma^{D}(e^{+}e^{-} \to \psi(3770) \to \pi^{0}J/\psi)$ and branching fraction $\mathcal{B}(\psi(3770) \to \pi^{0}J/\psi)$. Here ``N" indicates the corresponding uncertainty is negligible.}
    \setlength{\tabcolsep}{0.7mm}{
        \begin{tabular}{lcc}
            \hline \hline
            Source & $\sigma^{D}$  & $\mathcal{B}$  \\
            \hline
            Luminosity & 0.4 & 0.4 \\
            Tracking & 2.0 & 2.0 \\
            Photon detection & 2.0 & 2.0 \\
            Dep$_{\mu^\pm}$ criterion & 0.4 & 0.4 \\
            $M_{\gamma_{\text{H}}\mu^+\mu^-}$ requirement &  1.0 & 1.0 \\
            Kinematic fit & 0.4 & 0.4 \\
            Signal shape & 3.5 & 3.5 \\
            Flat background shape & N & N \\
            $\pi^0$ non-$J/\psi$ background & 1.8 & 1.8 \\
            $J/\psi$ non-$\pi^0$ background & N & N \\
            Fit range & N & N \\
            Input line shape & 3.2 & 3.2\\
            $\mathcal{B}(J/\psi \to \mu^{+}\mu^{-})$ & 0.6 & 0.6 \\
            $\mathcal{B}(\pi^{0} \to \gamma\gamma)$ & 0.1 & 0.1 \\
            $\sigma^D(\psi(3686) \to \pi^0 J/\psi)$ & 0.8 & 0.8 \\
             $\sigma^D(e^+e^- \to \psi(3770))$ & N & 7.8 \\
            \hline
            Total & 6.0 & 9.9 \\
            \hline \hline
        \end{tabular}
    }
    \label{Tab:uncertainty}
\end{table}

\begin{figure}[htbp]
\centering
\includegraphics[width=0.48\textwidth]{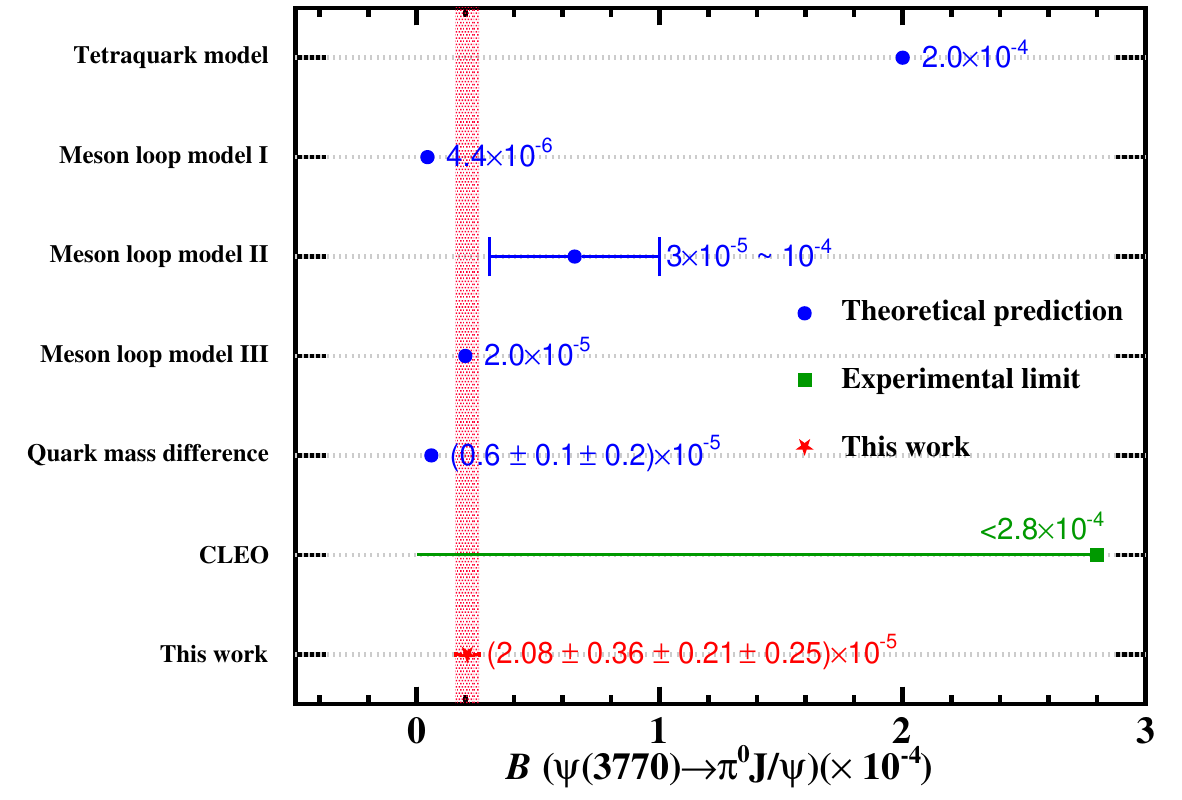}
\caption{Theoretical predictions including tetraquark model~\cite{Voloshin}, meson loop model I-III~\cite{predictpi0jpsi2, updatepi0jpsi3,Achasov:2005qb}, and the quark mass difference model~\cite{Ioffe:1980mx} , compared with previous experimental results and this measurement of $\mathcal{B}(\psi(3770)\to\pi^{0}J/\psi)$. The prediction from Ref.~\cite{Achasov:2005qb}(Meson loop model III) assumes $\psi(3770)$ has no isospin partner. For Ref.~\cite{Ioffe:1980mx}(Quark mass difference), the first uncertainty is propagated from the measured $\mathcal{B}(\psi(3770) \to \eta J/\psi)$, while the second represents theoretical uncertainties primarily from the characteristic mass scale of strong interactions.}
    \label{Fig:ratio}
\end{figure}

In summary, utilizing data collected at ~\mbox{$\sqrt{s} = 3.773$}~GeV with a total integrated luminosity of $20.3~\text{fb}^{-1}$, we have analyzed the process $e^+ e^- \to \pi^{0}J/\psi$. A signal with statistical significance of $6.1\sigma$ is observed for the first time, while the significance for $e^+ e^- \to \psi(3770) \to \pi^{0}J/\psi$ is determined to be $4.5\sigma$. The dressed cross section $\sigma^D(e^+ e^- \to \pi^{0}J/\psi)$ at $\sqrt{s} = 3.773$~GeV is measured as $(249 \pm 44 \pm 15)~\text{fb}$, and the branching fraction $\mathcal{B}(\psi(3770) \to \pi^{0}J/\psi)$ is determined to be $(2.08 \pm 0.36 \pm 0.21\pm 0.25) \times 10^{-5}$, where the first uncertainty is statistical, the second is systematic, and the third is the uncertainty arising from a possible interference with the $\psi(3686) \to \pi^{0}J/\psi$. This is the first measurement of this branching fraction, and is consistent with the previous best lower limit~\cite{CLEOresult}. This result allows for effective distinction between different theoretical calculations. The comparison of theoretical predictions and experimental measurements from CLEO and this work is shown in Fig.~\ref{Fig:ratio}. Our measured branching fraction is significantly lower than the prediction from Ref.~\cite{Voloshin}, which challenges the hypothesis of a substantial tetraquark component with vector isospin character within the $\psi(3770)$. Instead, it is in good agreement with theoretical calculations that incorporate meson loop effects~\cite{Achasov:2005qb}, underscoring the importance of these effects in this isospin-violating decay. 

Furthermore, utilizing the PDG values~\cite{PDG} and the $\mathcal{B}(\psi(3770)\to \pi^{0}J/\psi)$ measured in this work, we determine the ratio $\rm{R}^{\psi(3686)}/\rm{R}^{\psi(3770)} = 2.1\pm 1.3$. This ratio is consistent with the corresponding theoretical prediction $4.0$, derived under the pure $c\bar{c}$ state assumption for both $\psi(3686)$ and $\psi(3770)$~\cite{Ioffe:1980mx}, within two standard deviations. This suggests that, under the premise of a single resonance, the decay properties of $\psi(3770)$ are not in contradiction with the conventional charmonium model. Consequently, the $S$–$D$ mixing scheme, proposed to resolve the “$\rho-\pi$'' puzzle, remains a viable explanation.

However, a critical caveat must be noted regarding this interpretation. Our analysis treats the $\psi(3770)$ as a simple resonance (i.e., using the PDG-listed parameters based on the single-resonance  assumption as input). This may introduce systematic bias due to model oversimplification.  
In light of growing experimental evidence indicating that the  $\psi(3770)$ is likely a multi-resonance complex~\cite{BESIII:591}, the conclusions drawn under the single-resonance hypothesis require careful consideration. Therefore, more definitive conclusions await further precise measurements and dedicated theoretical investigations that explicitly account for the possible complex nature of this state.

The BESIII Collaboration thanks the staff of BEPCII (https://cstr.cn/31109.02.BEPC) and the IHEP computing center for their strong support. This work is supported in part by National Key R\&D Program of China under Contracts Nos. 2025YFA1613900, 2023YFA1606000, 
2023YFA1606704; National Natural Science Foundation of China (NSFC) under Contracts Nos. 12275058, 12035009, 11635010, 11935015, 11935016, 11935018, 12025502, 12035013, 12061131003, 12192260, 12192261, 12192262, 12192263, 12192264, 12192265, 12221005, 12225509, 12235017, 12342502, 12361141819; the Chinese Academy of Sciences (CAS) Large-Scale Scientific Facility Program; the Strategic Priority Research Program of Chinese Academy of Sciences under Contract No. XDA0480600; CAS under Contract No. YSBR-101; 100 Talents Program of CAS; The Institute of Nuclear and Particle Physics (INPAC) and Shanghai Key Laboratory for Particle Physics and Cosmology; ERC under Contract No. 758462; German Research Foundation DFG under Contract No. FOR5327; Istituto Nazionale di Fisica Nucleare, Italy; Knut and Alice Wallenberg Foundation under Contracts Nos. 2021.0174, 2021.0299, 2023.0315; Ministry of Development of Turkey under Contract No. DPT2006K-120470; National Research Foundation of Korea under Contract No. NRF-2022R1A2C1092335; National Science and Technology fund of Mongolia; Polish National Science Centre under Contract No. 2024/53/B/ST2/00975; STFC (United Kingdom); Swedish Research Council under Contract No. 2019.04595; U. S. Department of Energy under Contract No. DE-FG02-05ER41374

\bibliography{reference}

\begin{thebibliography}{41}
\bibitem [{sig()}]{sigma3686}%
  \BibitemOpen
  \href@noop {} {}\bibinfo {note} {The contribution from the $\psi(3686)$ tail
  at $\sqrt{s} = 3.773$ GeV will be subtracted directly when calculating
  $\sigma(\psi(3770)\to\pi^{0}J/\psi)$, and it is estimated by
  $\sigma^{tail}_{\psi_{3686}} = \left|
  \frac{\sqrt{12\pi\Gamma^{ee}_{\psi_{3686}}\Gamma^{\rm{tot}}_{\psi_{3686}}\mathcal{B}_{\pi^{0}J/\psi}}}{s-M_{\psi_{3686}}^{2}+i\Gamma^{\rm{tot}}_{\psi_{3686}}M_{\psi_{3686}}}
  \sqrt{\frac {PS(\sqrt{s})}{PS(M_{\psi_{3686}})}} \right|^{2}$, where
  $\mathcal{B}_{\pi^{0}J/\psi}$ represent the branching fraction of
  $\psi(3686)\to \pi^{0} J/\psi$, $PS$ is phase space factor which is defined
  as: $PS$($\sqrt{s}$)=q($\sqrt{s}$)$^{3}/s$,
  $PS(M_{\psi(3686)})$=$q(M_{\psi(3686)})^{3}/M_{\psi(3686)}^{2}$, $q(\sqrt s)
  = ((s-(M_{\pi^{0}} - M_{J/\psi})^{2})(s-(M_{\pi^{0}} +
  M_{J/\psi})^{2}))^{1/2}/2\sqrt s$, $q(M_{\psi(3686)}) =
  ((M_{\psi(3686)}^{2}-(M_{\pi^{0}} -
  M_{J/\psi})^{2})(M_{\psi(3686)}^{2}-(M_{\pi^{0}} +
  M_{J/\psi})^{2}))^{1/2}/2M_{\psi(3686)}$. All the parameters used in the
  formula are from PDG. The errors were calculated using the error propagation
  formula.}\BibitemShut {Stop}%
\end{thebibliography}
\end{document}